\UseRawInputEncoding
\documentclass[showpacs,superscriptaddress,twocolumn]{revtex4-2}

\usepackage{bm,graphicx,dcolumn,epstopdf,epsf,latexsym,mathbbol,dcolumn,amssymb,amsmath,color,slashed,mathrsfs,mathcomp,simplewick}

\usepackage{multirow}
\usepackage{tabularx}
\usepackage{makecell}
\usepackage{color, soul}
\usepackage{hyperref}
\usepackage{url}

\usepackage{amssymb}
\usepackage{amsmath}
\usepackage{amsfonts}
\usepackage{slashed}
\usepackage{graphicx}
\usepackage{relsize}
\usepackage{color}
\usepackage{extarrows}
\usepackage{cancel}
\usepackage{placeins}
\usepackage{float}
\usepackage{subcaption}
\usepackage{caption}
\usepackage{here}
\usepackage[normalem]{ulem}

\begin{document}

\title{
Constraints on Metric--Palatini Gravity from QPO Data}

\author{Elham Ghorani}
\email{elham.ghorani@sabanciuniv.edu}
\affiliation{Faculty of Engineering and Natural Sciences, Sabanc{\i} University, 34956 Tuzla, Istanbul, Turkey}

\author{Samik Mitra}
\email{m.samik@iitg.ac.in}
\affiliation{Department of Physics, Indian Insitute of Technology Guwahati, Assam, India, 781039}

\author{Javlon Rayimbaev}
\email{javlon@astrin.uz}
\affiliation{Institute of Fundamental and Applied Research, National Research University TIIAME, Kori Niyoziy 39, Tashkent 100000, Uzbekistan} 
\affiliation{Shahrisabz State Pedagogical Institute, Shahrisabz Str. 10, Shahrisabz 181301, Uzbekistan}
\affiliation{Tashkent State Technical University, Tashkent 100095, Uzbekistan}

\author{Beyhan Puli{\c c}e}
\email{beyhan.pulice@sabanciuniv.edu}
\affiliation{Faculty of Engineering and Natural Sciences, Sabanc{\i} University, 34956 Tuzla, Istanbul, Turkey}
\affiliation{Astrophysics Research Center,
The Open University of Israel, Raanana 4353701, Israel}

\author{Farruh Atamurotov}
\email{atamurotov@yahoo.com}
\affiliation{Inha University in Tashkent, Ziyolilar 9, Tashkent 100170, Uzbekistan}
\affiliation{Urgench State University, Kh. Alimdjan str. 14, Urgench 220100, Uzbekistan}

\author{Ahmadjon Abdujabbarov}
\email{ahmadjon@astrin.uz}
\affiliation{University of Tashkent for Applied Sciences, Str. Gavhar 1, Tashkent 100149, Uzbekistan}

\author{\fbox{Durmu{\c s} Demir}}
\email{durmus.demir@sabanciuniv.edu}
\affiliation{Faculty of Engineering and Natural Sciences, Sabanc{\i} University, 34956 Tuzla, Istanbul, Turkey \\ \textcolor{blue}{Dedicated to Durmu\c{s} Demir (1967-2024), our supervisor and candid friend.}}

\begin{abstract}
In this work, we study metric-Palatini gravity extended by the antisymmetric part of the affine curvature. This gravity theory leads to general relativity plus a geometric Proca field. Using our previous construction of its static spherically-symmetric AdS solution [Eur. Phys. J. C83 (2023) 4, 318], we perform a detailed analysis in this work using the observational quasiperiodic oscillations (QPOs) data. To this end, we use the latest data from stellar-mass black hole GRO J1655-40, intermediate-mass black hole in M82-X1, and the super-massive black hole in SgA* (our Milky Way) and perform a Monte-Carlo-Markov-Chain (MCMC) analysis to determine or bound the model parameters. Our results shed light on the allowed ranges of the Proca mass and other parameters. The results imply that our solutions can cover all three astrophysical black holes. Our analysis can also be extended to more general metric-affine gravity theories. 
\end{abstract}

\maketitle

\section{Introduction}

The ongoing research in astrophysics, gravitation, and cosmology concentrates on one single question: Is general relativity (GR) the sole theory of gravitation? The answer to this question requires a detailed study of physically plausible extensions of the GR. One such extension concerns Riemannian geometries in which the metric and connections remain independent geometrical quantities \cite{schroedinger,mag1,Vitagliano2011}. The simplest example of such an extension is metric-Palatini gravity \cite{harko2012,Capozziello2013a,Capozziello2015}, which has been investigated in the contexts of dark matter \cite{Capozziello2012a}, wormholes \cite{Capozziello2012b}, and cosmology \cite{Capozziello2012c}. The metric-Palatini gravity in which the non-metricity tensor gives rise to a geometric $Z^\prime$ field that forms a special class \cite{Demir2020}, and it has been studied regarding gravitational waves \cite{Shaaban}, and black hole properties in the Schwarzschild \cite{dp-yeni} and AdS \cite{AdS-Proca-1} backgrounds. The present work is a sequel to our previous work \cite{AdS-Proca-1} and aims at constraining the metric-Palatini gravity with the QPO data.  

The metric-Palatini gravity has been discussed in detail in \cite{dp-yeni,AdS-Proca-1}. Here, we want to discuss its salient features briefly. It  is characterized by a metric $g_{\mu\nu}$ and a torsion-free affine connection $\Gamma^\lambda_{\mu\nu}$, which is  independent of the Levi-Civita connection ${}^g\Gamma^{\lambda}_{\mu\nu}$ of the metric. The connection can be decomposed into geometrical scalars, vectors, and tensors \cite{Demir2012}, where the geometrical vector is nothing but the non-metricity vector. Metric-Palatini gravity has rather widespread effects, such as the symmergent gravity restoring gauge symmetry \cite{Demir2019, Demir2021, Demir2023}, natural inflation \cite{bauer-demir1,bauer-demir2}, and astrophysical and cosmological impacts \cite{Palatini-f(R),2024PDU....4601577K,2024PDU....4501514Z} of higher-curvature terms \cite{Vitagliano2011,Vitagliano2013,Demir2020}.

Addition of a term like ${\mathbb{R}}_{[\mu\nu]}(\Gamma) {\mathbb{R}}^{[\mu\nu]}(\Gamma)$, where ${\mathbb{R}}_{[\mu\nu]}(\Gamma)$ is the antisymmetric part of the affine Ricci tensor ${\mathbb{R}}_{\mu\nu}(\Gamma)$, takes the 
 Palatini formulation is one step further. This addition is special and important because it gives rise to the GR plus a massive geometric vector field $Q_{\mu}$ \cite{Vitagliano2010,Demir2020}. This vector field, a geometric Proca field, is defined as 
 $Q_\mu \equiv \frac{1}{4} Q _{\mu \nu}^{~~~\nu}$ where 
 $Q_{\lambda\mu\nu}\equiv -{}^\Gamma \nabla_\lambda g_{\mu\nu}$ is the non-metricity tensor \cite{Demir2020, Buchdahl1979,Tucker1996,Obukhov1997,Vitagliano2010}. Without torsion, this non-metricate vector is the only source of deviations from the GR.  Geometric Proca is a direct signature of metric-incompatible symmetric connections (torsion-free).  It is not something put by hand. It is not a gauge field; rather, it is a geometrical massive vector field \cite{Demir2020} having specific coupling patterns to quarks and leptons \cite{dp-yeni}. This Palatini formalism can be further formulated by including in the action the metrical curvature $R_{\mu\nu}({}^g\Gamma)$ in addition to the affine curvature ${\mathbb{R}}_{\mu\nu}(\Gamma)$. Except for the quadratic term ${\mathbb{R}}_{[\mu\nu]}(\Gamma) {\mathbb{R}}^{[\mu\nu]}(\Gamma)$ leading to the geometric Proca field, this combined metric-affine framework leads to the metric-Palatini gravity \cite{harko2012,Capozziello2013a,Capozziello2015}. The gravity theory we study in this work is nothing but the metric-Palatini gravity extended with the ${\mathbb{R}}_{[\mu\nu]}(\Gamma) {\mathbb{R}}^{[\mu\nu]}(\Gamma)$ invariant and a negative cosmological constant (CC) \cite{AdS-Proca-1}.  Indeed, as was revealed in \cite{dp-yeni}, in the presence of the geometric Proca $Q_{\mu}$, the CC is a prerequisite for static spherically symmetric solutions. We call our framework 
 extended metric-Palatini gravity (EMPG). Its action takes the schematic form (given in \cite{AdS-Proca-1})  
\begin{eqnarray}
\label{EMPG}
S[g,\Gamma]=\!\!\int\!\! d^4x \sqrt{-g} &&\Bigg \{\!``g^{\mu\nu}{R}_{\mu\nu}\left({}^g\Gamma\right)" + ``g^{\mu\nu}{\mathbb{R}}_{\mu\nu}\left(\Gamma\right)"\nonumber\\
&&+ ``{\mathbb{R}}_{[\mu\nu]}(\Gamma) 
 {\mathbb{R}}^{[\mu\nu]}(\Gamma)" + ``{\rm CC}" \!\Bigg \}
\end{eqnarray} 
which is an Einstein-geometric Proca-Anti de Sitter (AdS) gravity theory. Its geometric origin differentiates it from the Einstein-Proca systems in the literature, which have been analyzed  for finding Reissner-Nordstr\"{o}m type spherically-symmetric vacuum solutions  \cite{Tresguerres1995a, 
Tucker1995, Vlachynsky1996, Macias1999},  for determining the role of the Proca field \cite{Bekenstein1971, Bekenstein1972, Adler1978}, for obtaining static spherically symmetric solutions \cite{Frolov1978, Gottlieb1984, Leaute1985}, and for revealing the structure of the horizon radius \cite{Ayon1999, Obukhov1999, Toussaint2000}. Our goal in the present work is to extend our previous work \cite{AdS-Proca-1} by determining constraints on 
the EMPG parameters from the QPOs. 

Analyzing the spectroscopic properties of radiation released by accreting matter in the neighborhood of gravitational compact objects, such as black holes, wormholes, and neutron stars, is one possible way to investigate the properties of the spacetime around them. The radiation processes around the accretion disk are shaped mainly by the gravitational pull of the black hole in binary systems where neutron stars or black holes coexist with their companion stars. The continuous X-ray data from these accretion disks in (micro)quasars can be Fourier-analyzed \cite{Stella1998ApJL,Stella1999ApJ} to
reveal the existence of QPOs (quasi-periodic oscillations). The QPOs are categorized as high frequency (HF) when their peak frequencies fall within the range of approximately 0.1 to 1 kilohertz (kHz) and as low frequency (LF) when their frequencies are below about 0.1 kHz \cite{Stuchlik2000AA}.

Accurate measurements of QPO frequencies in quasars and binary systems (microquasars) have shed light on the physical processes driving their formation. This ongoing study attempts to determine the properties of the accretion disk's inner edge and test different theories of gravity in this way. These tests can potentially yield measurements concerning the black hole properties and the radii of nearby ISCOs. Previous research has demonstrated that investigating the QPO orbit can be instrumental in inferring the radius of the ISCO lying close to the QPO orbit \cite{Rayimbaev2021Galax...9...75R,Rayimbaev2021QPO,Rayimbaev2022CQGra,Rayimbaev2022EPJCEMSQPO,Rayimbaev2022IJMPDQPOcharged,Rayimbaev2022PDU,2021Galax...9...65T,2021Univ....7..307A,2024PDU....4401483R,2023Galax..11...95R,2023Univ....9..391M,Rayimbaev2023AnPhy.45469335R,Rayimbaev2023EPJC...83..572R,Rayimbaev2023EPJC...83..730Q,2022Univ....8..507T,2024PhyS...99f5011A,2021IJMPD..3050037T,2023Galax..11...70T,2024JHEAp..43...51D,FENG2024158}. Notably, the relativistic precision model indicates that the distance between these orbits is within the measurement error range.

In the present work, we probe and constrain the EMPG model regarding its Einstein-Geometric Proca-AdS compact object using the existing quasiperiodic oscillations (QPO) data. Our work consists of two parts: (i) The description of the gravity model and corresponding spherically symmetric solution, and (ii) the observational implications of the solutions.  The paper is organized as follows. Sect.~\ref{sect:2} is devoted to reviewing EMPG \cite{AdS-Proca-1} and its spherically symmetric solution. Particle dynamics around Einstein-Geometric Proca-AdS compact object will be studied in Sect.~\ref{sect:3}. The fundamental frequencies of circular orbits will be discussed in Sect.~\ref{sect:4}. We investigate QPOs and the corresponding astrophysical applications in Sect.~\ref{sect:5}. In Sect.~\ref{sect:6}, we determine the constraints of the EMPG using observational data from the twin peak QPOs.  We conclude in Sect.~\ref{sect:7}.  

\section{Static Spherically-Symmetric Solutions in EMPG Model\label{sect:2}}

In this section, we give a detailed discussion of the EMPG model. The material here has been largely contained in our previous work \cite{AdS-Proca-1}, and we summarize the main results. The EMPG action is given by \cite{Demir2020,dp-yeni,AdS-Proca-1}
 \begin{eqnarray}\nonumber
S[g,\Gamma]&=&\int d^4x \sqrt{-g} \Bigg \{
\frac{M^2}{2} {R}\left(g\right) + \frac{{\overline{M}}^2}{2} {\mathbb{R}}\left(g,\Gamma\right) \\
&+& \xi {\overline{\mathbb{R}}}_{\mu\nu}\left(\Gamma\right) {\overline{\mathbb{R}}}^{\mu\nu}\left(\Gamma\right) -V_0 + 
{\mathcal{L}}_{m}({}^g\Gamma,\psi) \Bigg \}, 
\label{mag-action}
\end{eqnarray}
where the affine curvatures in this action follow from the affine Riemann curvature
\begin{eqnarray}
\label{affine-Riemann}
{\mathbb{R}}^\mu_{\alpha\nu\beta}\left(\Gamma\right) = \partial_\nu \Gamma^\mu_{\beta\alpha} - \partial_\beta \Gamma^\mu_{\nu\alpha} + \Gamma^\mu_{\nu\lambda} \Gamma^\lambda_{\beta\alpha} -\Gamma^\mu_{\beta\lambda} \Gamma^\lambda_{\nu\alpha}
\end{eqnarray}
 with the obvious property ${\mathbb{R}}^\mu_{\alpha\nu\beta}\left(\Gamma\right)=-{\mathbb{R}}^\mu_{\alpha\beta\nu}\left(\Gamma\right)$. Its contractions give rise to two distinct affine Ricci tensors: the canonical tensor ${\mathbb{R}}_{\mu\nu}\left(\Gamma\right) = {\mathbb{R}}^\lambda_{\mu\lambda\nu}\left(\Gamma\right)$, and the antisymmetric Ricci tensor ${\overline{\mathbb{R}}_{\mu\nu}}\left(\Gamma\right) = {\mathbb{R}}^\lambda_{\lambda\mu\nu}\left(\Gamma\right) = {\mathbb{R}}_{[\mu\nu]}\left(\Gamma\right)$.
The latter vanishes
identically in the metrical geometry, ${\overline{\mathbb{R}}_{\mu\nu}}\left({}^g\Gamma\right) = 0$.
The term proportional to $M^2$ in the action (\ref{mag-action}) corresponds to the Einstein-Hilbert term in GR. The term proportional to ${\overline{M}}^2$ corresponds to the linear case of metric-Palatini gravity. The third term, proportional to $\xi$, gives the extension of metric-Palatini gravity with the antisymmetric part of the affine Ricci curvature \cite{Demir2020,dp-yeni}. In the last two terms, we separate the vacuum energy density $V_0$ from the Lagrangian of matter ${\mathcal{L}}_{m}({}^g\Gamma,\psi) $ that governs the dynamics of the matter fields $\psi$. 

In the presence of the Levi-Civita connection ${}^g\Gamma^\lambda_{\mu\nu}$, the torsion-free affine connection can always be decomposed as 
\begin{align}\Gamma^\lambda_{\mu\nu}= {}^g\Gamma^\lambda_{\mu\nu} +  \frac{1}{2} g^{\lambda \rho} ( Q_{\mu \nu \rho } + Q_{\nu \mu \rho } - Q_{\rho \mu \nu} ),
\label{fark-connection}
\end{align}
where $Q_{\lambda \mu \nu} = - {}^\Gamma \nabla_{\lambda} g_{\mu \nu}$ is the non-metricity tensor. The use of this decomposition in the metric-Palatini action (\ref{mag-action}) leads to the reduced action \cite{Demir2020, dp-yeni,AdS-Proca-1}
\begin{eqnarray}\label{action-reduced}
S[g,Y,\psi] &=& \int d^4 x \sqrt{-g} \Bigg\{\frac{1}{16 \pi G_N} R(g)-V_0 \\ \nonumber &-& \frac{1}{4} Y_{\mu \nu} Y^{\mu \nu} - \frac{1}{2} M_Y^2 Y_{\mu} Y^{\mu}+{\mathcal{L}}_{m} (g,{}^g \Gamma,\psi) \Bigg \}
\end{eqnarray}
in which $Q_\mu = Q _{\mu \nu}^{\nu}/4$ is the non-metricity vector,  $Y_\mu=2 \sqrt{\xi} Q_\mu$ is the canonical geometric Proca field, $G_N=8\pi/(M^2 + \overline{M}^2)$ is Newton's gravitational constant, and
\begin{eqnarray}
M_Y^2 = \frac{3 \overline{M}^2 }{2 \xi}
\end{eqnarray}
is the squared mass of the $Y_\mu$. For analyses in this work, it is convenient to write the reduced action (\ref{action-reduced}) in geometrical units as 
 \begin{eqnarray}\label{action-EGP}
     S[g,Y] &=& \int d^4 x \sqrt{-g} \frac{1}{2 \kappa} \Bigg\{R(g) - 2 \Lambda \\ \nonumber &-&  M_Y^2 \hat{Y}_{\mu} \hat{Y}^{\mu}
- \frac{1}{2} \hat{Y}_{\mu \nu} \hat{Y}^{\mu \nu} \Bigg\}
 \end{eqnarray}
in which $\kappa = 8 \pi G_N$, $\Lambda=8\pi G_N V_0$ is the CC, and $\hat{Y}_\mu \equiv \sqrt{\kappa} Y_{\mu}$ is the canonical dimensionless Proca field. From this action, the motion equations for $g_{\mu\nu}$ and $\hat{Y}_\mu$ are found to be
\begin{eqnarray}\label{Einstein-eqns}\nonumber
   R_{\mu \nu} - \Lambda g_{\mu \nu} - \hat{Y}_{\alpha\mu} \hat{Y}^{\alpha}_{\;\;\;\nu}  + \frac{1}{4} \hat{Y}_{\alpha \beta} \hat{Y}^{\alpha \beta} g_{\mu \nu} - M_Y^2 \hat{Y}_{\mu} \hat{Y}_{\nu} = 0, 
\end{eqnarray}
and 
\begin{align}
\nabla_\mu \hat{Y}^{\mu \nu} - M^2_Y \hat{Y}^\nu = 0.
\label{eom-Y}
\end{align}
These equations have been analyzed in \cite{dp-yeni} and \cite{AdS-Proca-1} to find black hole solutions with $\Lambda=0$ and $\Lambda<0$, respectively. In search for a general spherically-symmetric and static solution, the ansatz
\begin{align}
g_{\mu \nu} = \text{diag}(-h(r),\frac{1}{f(r)} ,r^2, r^2 \sin ^2 \theta).
\label{metric-sss}
\end{align}
and 
\begin{align}
\hat{Y}_\mu = \hat{\phi}(r) \delta_\mu^0. 
\label{proca-sss}
\end{align} 
lead to the solution
\begin{equation}
\hat{\phi}(\hat{r})=\frac{q_1}{\hat{r}^{\frac{1-\sigma}{2}}} + \frac{q_2}{\hat{r}^{\frac{1+\sigma}{2}}} 
\label{psi}
\end{equation}
with the mass parameter
\begin{align}\label{sigma}
\sigma=\sqrt{1 + 4 \hat{M}_Y^2 l^2} ~    
\end{align}
where $l$ stands for the AdS radius.
The Breitenlohner-Freedman mass bound \cite{bf1,bf2} 
\begin{align}
0 \leq \sigma < 1 
\label{sigma-range}
\end{align}
prevents tachyonic run-away instabilities in the AdS background, with $q_1$ and $q_2$ interpreted as uniform potential and electromagnetic-like charge, respectively. Corresponding to the geometric Proca solution in (\ref{psi}), the metric potential $f$ and $h$ take the form \cite{AdS-Proca-1}
\begin{align}
\label{metric-funcs-param}
f(\hat{r}) &= \hat{r}^2l^{-2} + 1 + \frac{n_1}{\hat{r}^{1 - \sigma}} + \frac{n_2}{\hat{r}}\ ,\nonumber\\
h(\hat{r}) &= \hat{r}^2l^{-2} + 1 +\frac{m_1}{\hat{r}^{1 - \sigma}}+\frac{m_2}{\hat{r}}\ ,
\end{align}
in which
\begin{align}
\label{solution}
&n_1 = \frac{1 - \sigma}{4} q_1^2\ , \quad m_1 = \frac{1 - \sigma}{3 - \sigma} q_1^2\ , \nonumber \\
&n_2 = m_2 - \frac{(1 - \sigma)(1 + \sigma)}{6} q_1 q_2 \ , 
\end{align}
By setting $q_1=0$, we will recover Ads-Schwarzschild solution.
The ADM mass of this compact object takes the form \cite{AdS-Proca-1}
\begin{eqnarray}
\label{mass}
    M_{ADM}=\frac{1}{2}\left(q_1 q_2 \left[\gamma  \sigma +\frac{1}{3} (1-\sigma ) (\sigma +4)\right]-m_2\right)
\end{eqnarray}
in which $\gamma$ is the coefficient of the surface term for the geometric Proca 
after the normalization $M_{ADM}=1$. All these physical properties of the Einstein-geometric Proca AdS solution, including its horizon radius behavior with respect to the model parameters and the singularity structure, have already been analyzed in \cite{AdS-Proca-1}.

\section{Particle Motion Around EGP AdS Compact Objects \label{sect:3}}

In this section, we study the motion of test particles around the EMPG compact object solution above. The Lagrangian for such a particle reads as
\begin{equation}
    L_p=\frac{1}{2}mg_{\alpha\beta}\dot x^{\alpha}\dot x^{\beta} , \label{lag}
\end{equation}
in which $m$ is the mass of the particle. One finds the conservation laws  
\begin{align}
    g_{tt} \dot t= -\frac{E}{m} , \quad
    g_{\phi\phi} \dot \phi= \frac{L}{m} , 
    \label{conserved}
\end{align}
in terms of the conserved energy $E$ and conserved angular momentum $L$ of the particle. With the normalization
\begin{equation}
    g_{tt}\dot t^{2}+g_{rr}\dot r^{2}+g_{\theta\theta}\dot\theta^{2}+g_{\phi\phi}\dot \phi^{2}=-1 ,
   \label{norm}
\end{equation}
we get
\begin{equation}
    g_{rr}\dot r^{2}+g_{\theta\theta}\dot\theta^{2}=-\left(1+\frac{E^2}{m^2g_{tt}}+\frac{L^2}{m^2g_{\phi\phi}}\right) ,
    \label{norm2}
\end{equation}
by using the conserved quantities in (\ref{conserved}). We can rearrange (\ref{norm2}) as
\begin{equation}
    g_{rr}\dot r^{2}+g_{\theta\theta}\dot\theta^{2}=-\Pi(r,\theta) 
    \label{norm3}    
\end{equation}
with
\begin{equation} 
    \Pi(r,\theta)=1+\frac{\mathcal{E}^2}{g_{tt}}+\frac{\mathcal{L}^2}{g_{\phi\phi}}
    \label{pi}
\end{equation}
where $\mathcal{E}$ and $\mathcal{L}$ are $-E/m$ and $L/m$, respectively. In the azimuthal $\theta=\pi/2$ plane, the motion in the radial direction obeys the equation 
\begin{equation}
    g_{rr}\dot r^{2}=-\Pi(r,\theta_0)=R(r)
    \label{rnorm}
\end{equation}
where
\begin{equation}
R(r)=-\frac{1}{g_{tt}}(\mathcal{E}^2-V_{\rm eff}(r)) ,
\end{equation}
with the effective potential
\begin{eqnarray}
\nonumber
V_{\rm eff}(r)=-g_{tt}\left(1+\frac{\mathcal{L}^2}{g_{\phi\phi}}\right) = h(r)\left(1+\frac{\mathcal{L}^2}{r^2}\right) .
\end{eqnarray}

\begin{figure}[ht!]
 \centering
    \includegraphics[scale=0.58]{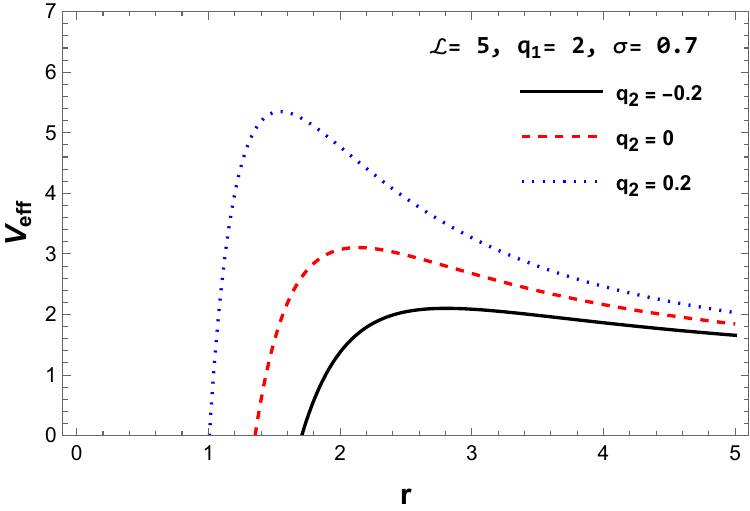}
   \includegraphics[scale=0.58]{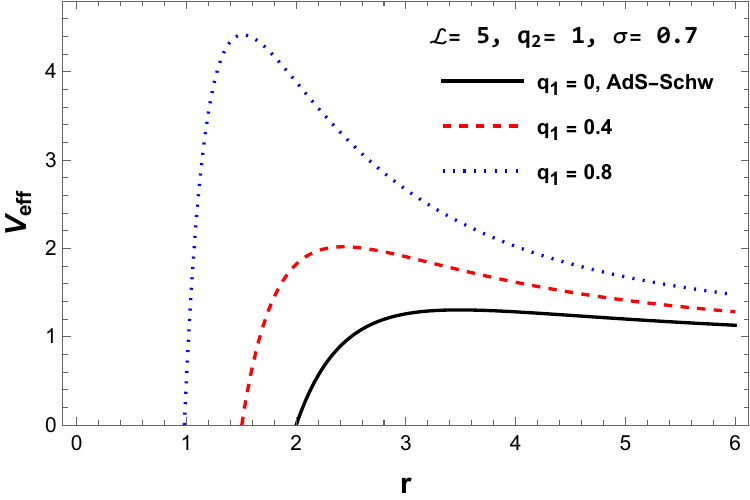}
 \caption{Variation of the effective potential $V_{\rm eff}$ with the radial coordinate $r$ for the indicated parameter values. In the upper (lower) panel, $q_1$ ($q_2$) is fixed. }\label{potentialplot}
\end{figure}

\begin{figure*}[ht!]
 \centering
   \includegraphics[scale=0.58]{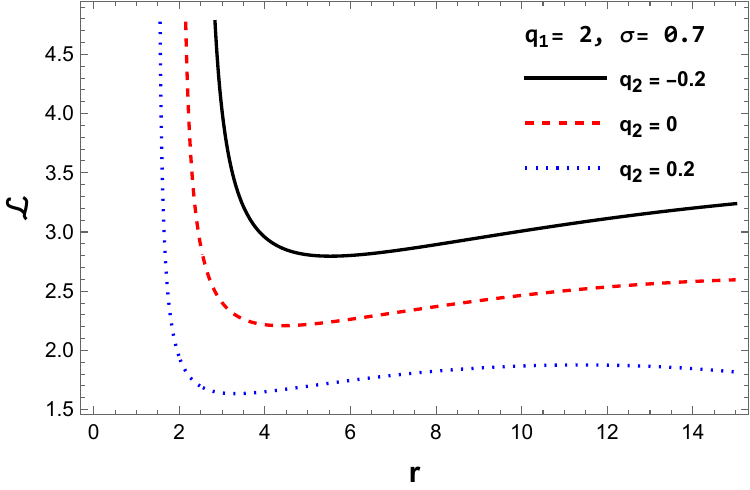}
   \includegraphics[scale=0.57]{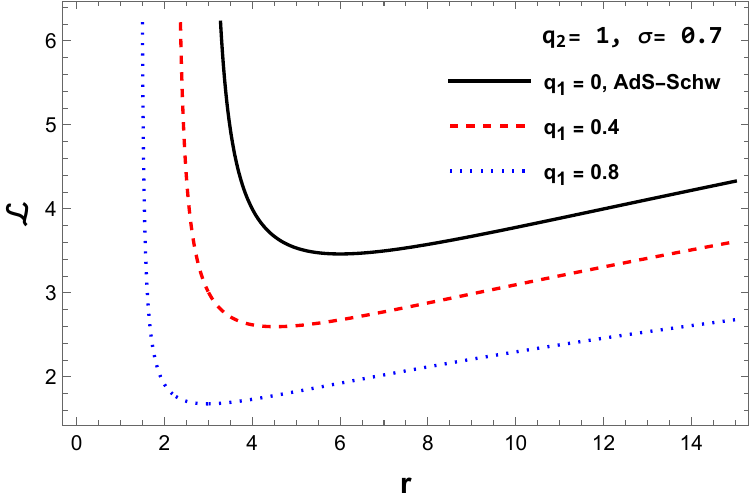}
   \includegraphics[scale=0.58]{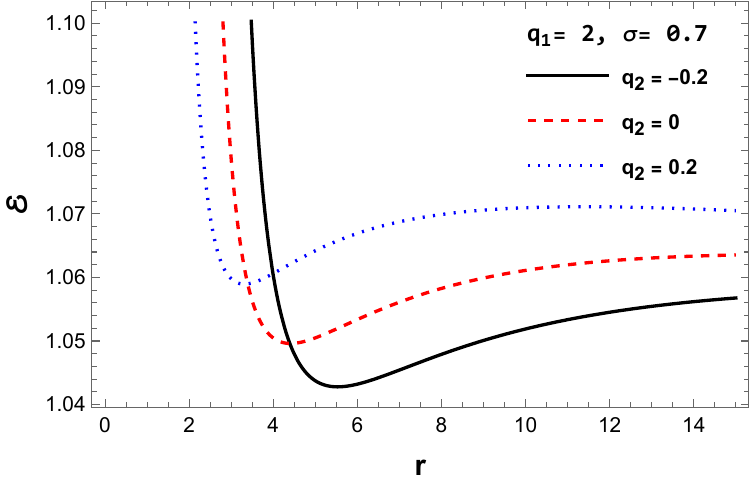}
   \includegraphics[scale=0.58]{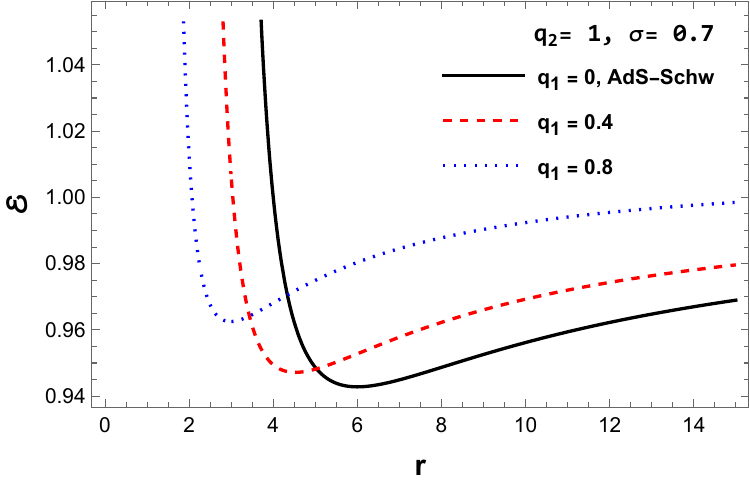}
 \caption{Variations of the angular momentum ${\mathcal{L}}$ and energy ${\mathcal{E}}$ with the radial coordinate $r$ for circular orbits for the indicated values of the parameters. In the left (right) panels, $q_{1}$ ($q_2$) is held fixed while $q_2$ ($q_1$)  takes on three different values.}\label{LE}
\end{figure*}

\begin{figure*}[ht!]
 \centering
   \includegraphics[scale=0.58]{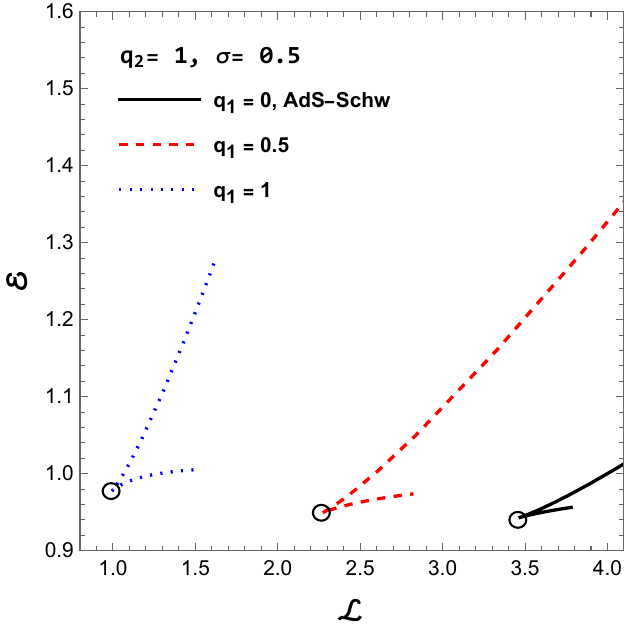}
   \includegraphics[scale=0.58]{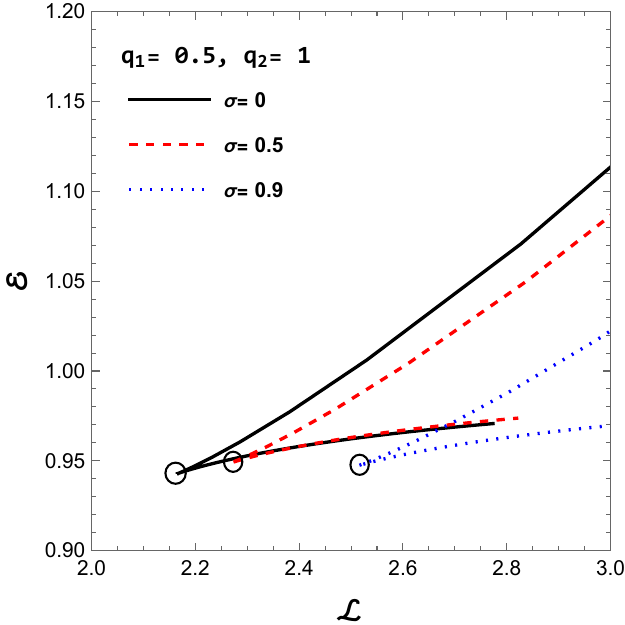}
 \caption{The energy ${\mathcal{E}}$ of the test particle around the EMPG compact object in circular orbits for specific ranges of the angular momentum ${\mathcal{L}}$. In the left (right) panel, $q_{2}$ and $\sigma$ are held fixed while $q_1$ varies. In the right panel, $q_1$ and $q_2$ are fixed with $\sigma$ taking on different values. }\label{cr}
\end{figure*}

The behavior of the effective potential is depicted in Fig. \ref{potentialplot}. The radius at which $V_{\rm eff}$ is zero corresponds to the horizon. According to the 
two panels, the horizon radius decreases while $q_1$ or $q_2$ increases. The case of $q_1=0$ corresponds to the AdS-Schwarzschild case.
If, on the other hand, the particle is at a circular orbit at $r = r_0$ with $\theta$ varying, then the equation of motion (\ref{norm3}) takes the form
\begin{equation}
    g_{\theta\theta}\dot \theta^{2}=\Theta(\theta)=-\Pi(r_0,\theta)
    \label{tnorm}
\end{equation}
such that the conditions for a particle to have a circular orbit at $r=r_0$ and $\theta=\theta_0$ read as 
\begin{equation}
     R(r_0)=0 , \
     \left.\frac{d R(r)}{dr}\right|_{r_{0}}=0
     \label{radial}
\end{equation}
which can be shown to be equivalent to the radial and polar conditions 
\begin{equation}
    \mathcal{E}^2=V_{\rm eff}(r_0) , \
    \left.\frac{d R(r)}{dr}\right|_{r_{0}}=0\,,
    \label{radial2}
\end{equation}
and
\begin{equation}
    \Theta(\theta_0)=0 , \  \left.\frac{d \Theta(\theta)}{d\theta}\right|_{\theta_{0}}=0\,.
    \label{theta}
\end{equation}
This polar condition on $\Theta(\theta)$ leads to 
zero angular momentum $\mathcal{L}=0$, which means that there are no off-equatorial circular orbits. In the radial equations ($\ref{radial}$) and ($\ref{radial2}$), one can determine the energy ${\mathcal{E}}$ and the angular momentum of circular orbits. Angular momentum (energy) is plotted in  Fig. \ref{LE} in the upper (lower) two panels for different values of $q_1$ and $q_2$. When $q_1$ or $q_2$ increases, the angular momentum decreases while keeping one of them held fixed in each panel. The opposite is true for the energy. Nevertheless, in all cases, the energy and angular momentum minima are shifted towards the smaller radii as $q_1$ or $q_2$ increases.  In Fig. \ref{cr} depicted are the allowed values of the energy and angular momentum of the particle in the circular orbits. For a fixed value of the angular momentum, the particle's energy ranges between a minimum and maximum. When they coincide, the angular momentum reaches its critic values, circled in Fig. \ref{cr}. As $q_1$ increases, the critical value of angular momentum gets smaller. However, when both $q_1$ and $q_2$ are fixed, the critical angular momentum takes larger values as the Breitenlohner-Freedman parameter  $\sigma$ increases.

\begin{figure*}[ht!]
 \centering
    \includegraphics[scale=0.58]{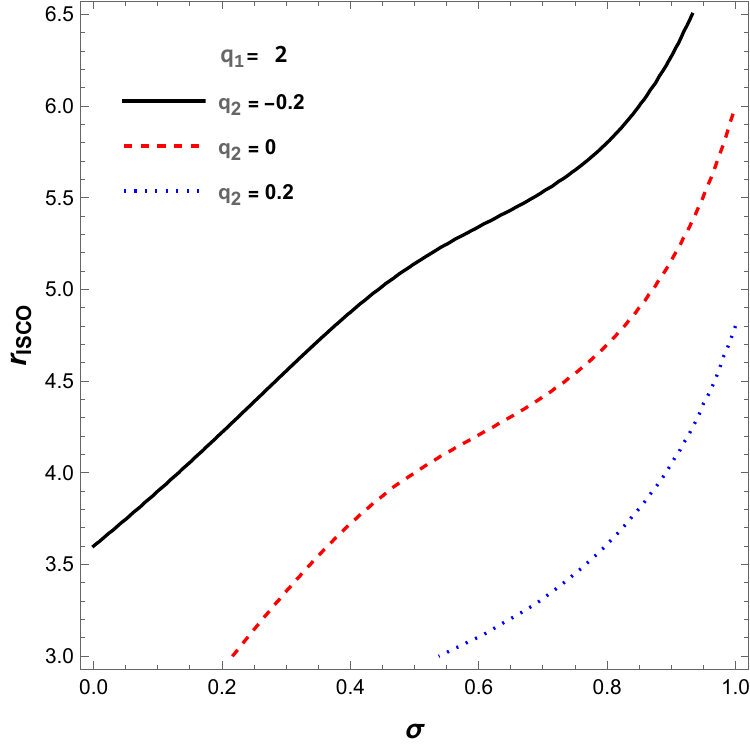}
   \includegraphics[scale=0.58]{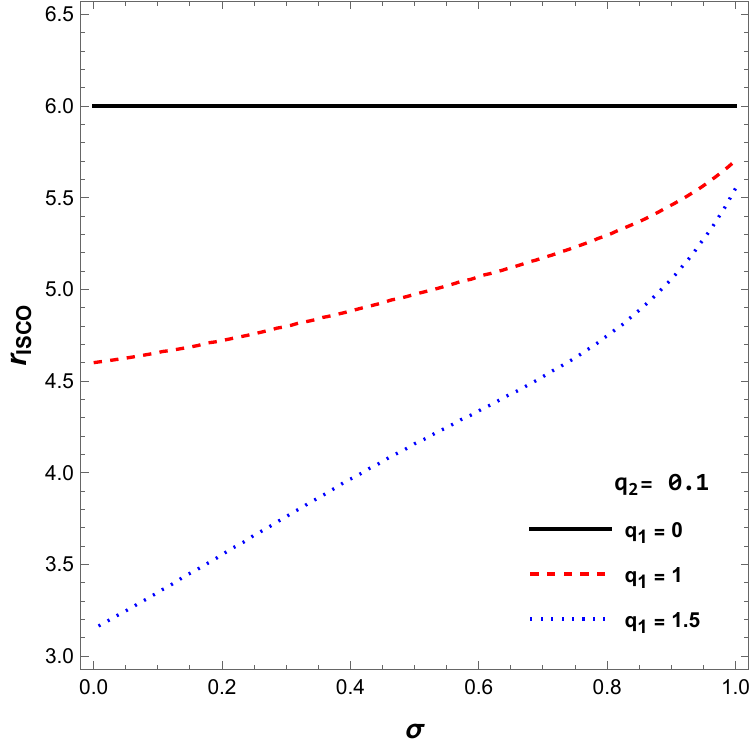}
 \caption{Dependence of the ISCO radius on the mass parameter $\sigma$ for the indicated values of charges. In the left (right) panel, $q_{1}$ ($q_2$) is fixed and $q_{2}$ ($q_1$) takes on three different values.}\label{iscoplot}
\end{figure*}

In addition to the radial conditions in \ref{radial2}, the following condition should be satisfied 
\begin{equation}
\left.\frac{d^2 V_{\rm eff}(r)}{dr^2}\right|_{r_{0}}=0
\label{radial3}
\end{equation}
for determining the ISCO radius. We solve this equation using the energy and angular momentum values we found from (\ref{radial2}). The resulting ISCO radii are shown in Fig. \ref{iscoplot}. As the figure suggests, the ISCO radius decreases as $q_1$ or $q_2$ increases, while the other remains fixed. However, as $\sigma$ increases, the ISCO radius increases.  For $q_1=0$, the ISCO radius remains constant and is almost equal to the Schwarzschild ISCO radius, as expected. 

\begin{figure*}[ht!]
 \centering
   \includegraphics[scale=0.58]{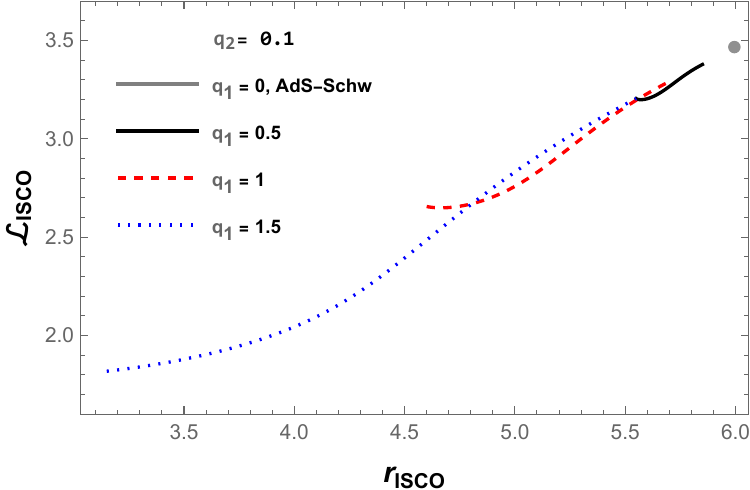}
   \includegraphics[scale=0.58]{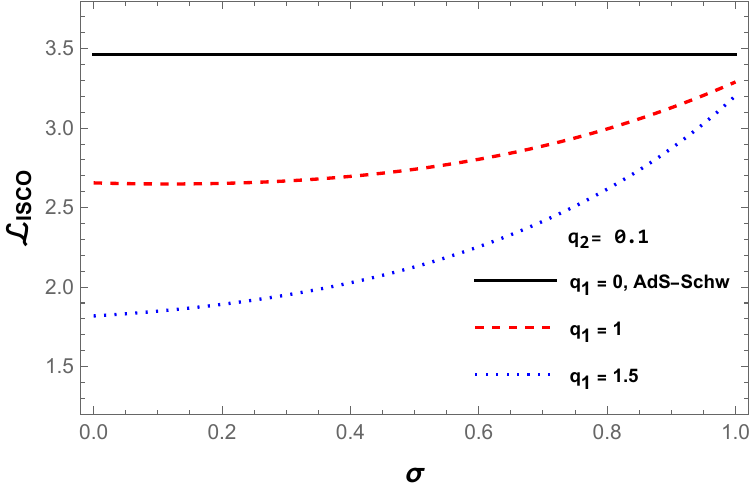}
   \includegraphics[scale=0.58]{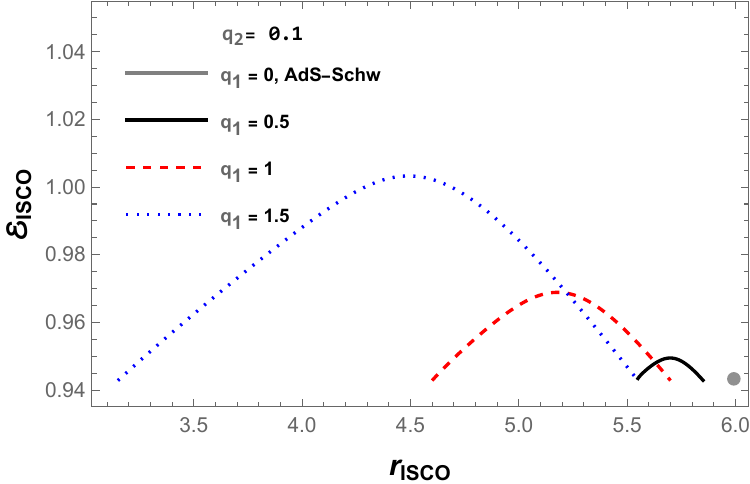}
   \includegraphics[scale=0.58]{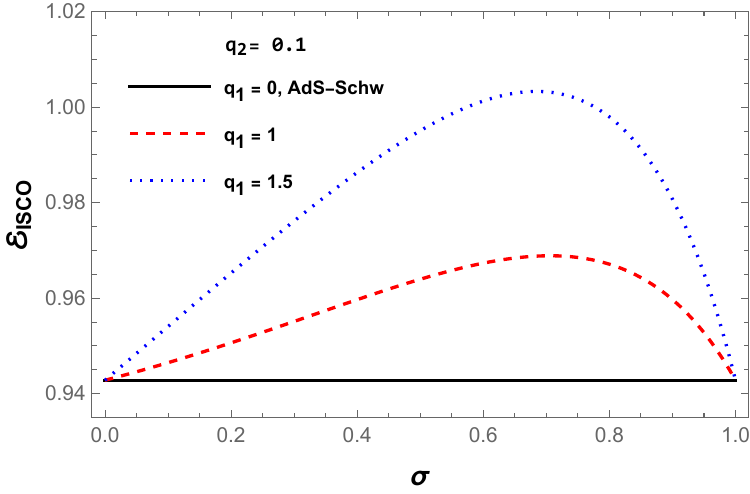}
 \caption{The dependencies of the energy and angular momentum of a particle (situated on ISCO)  on the ISCO radius (left panels) and on the $\sigma$ (right panels) for the indicated values of the Proca field parameters.}\label{LE2}
\end{figure*}

Fig. \ref{LE2} illustrates how the energy and angular momentum of the test particles vary with the ISCO radius. For $q_1=0$, the angular momentum takes its maximum value while energy takes its minimum value, corresponding to the Schwarzschild black hole with $r_{ISCO}=6$.  As $q_1$ increases, the ISCO radius also increases; therefore, the accessible energy ranges and angular momentum get larger.  For non-zero $q_1$, angular momentum increases with $\sigma$ and gets close to the Schwarzschild case as $\sigma$ tends to unity. The particle attains its minimum energy for $q_1 \neq 0$, corresponding to the Schwarzschild case at $\sigma=0$ and $\sigma=1$ values. The energy peak is shifted to higher values of $\sigma$ as $q_1$ increases.

\section{Fundamental frequencies\label{sect:4}}

This section calculates the fundamental frequencies characterizing the particle orbiting around the AdS geometric-Proca compact object. In particular, we focus on frequencies of Keplerian orbits and the radial and vertical oscillations. In invariant time $\lambda$ (namely $\dot{t}=dt/d\lambda$), the  angular velocity $\Omega_k=\Dot{\phi}/\Dot{t}$ of the particle,
\begin{equation}
    \Omega_k= -\frac{g_{tt}\mathcal{L}}{g_{\phi\phi}\mathcal{E}}
\end{equation}
which is nothing but the Keplerian frequency. Substituting energy, angular momentum, and metric functions leads to the following equation for Keplerian frequency:
\begin{equation}
    \Omega_k^2= \frac{1}{l^2}+\frac{q_1^2 (\sigma -1)^2 r^{\sigma -3}}{2 (\sigma -3)}+\frac{q_1 q_2 \left(\sigma ^2-4\right)+6}{6 r^3}.
\end{equation}
 we study small oscillation frequencies of the test particles to characterize their deviations from the stable circular orbit at $r=r_0$ and $\theta=\theta_0$. In the equatorial plane, motion equations in the radial direction read as 
\begin{equation}  \label{rad}
    g_{rr}\dot r^{2}=R(r)\ , \quad
    R(r_0)=0 \ , \quad
    \left.\frac{d R(r)}{dr}\right|_{r_{0}}=0 \, .
\end{equation}
Similarly, at a fixed radius, motion equations in the polar direction read as 
\begin{equation} \label{tet}
   g_{\theta\theta}\dot \theta^{2}=\Theta(\theta)\ , \quad
    \Theta(\theta_0)=0 \ , \quad
    \left.\frac{d \Theta(\theta)}{d\theta}\right|_{\theta_{0}}=0\ .
\end{equation}
Now, around an orbit of test particles defined by equations (\ref{rad}) and (\ref{tet}), we introduce the perturbations $\delta r$ and $\delta \theta$ as
\begin{equation}
        r=r_0+\delta r\nonumber\ , \quad 
    \theta=\theta_0+\delta \theta
\end{equation}
and study the time development of the perturbations to determine if the orbits are stable. In fact, 
we can expand $R(r)$ and $\Theta(\theta)$ around the stable circular orbit $(r_0,\theta_0)$ in the form  
\begin{align}
    &R(r_0+\delta r)=R(r_0)+R^\prime(r_0)\delta r+\frac{1}{2}R^{\prime\prime} (r_0)\delta r^2+... \, ,\nonumber \\ 
    &\Theta(\theta_0+\delta \theta)=\Theta(\theta_0)+\Theta^\prime(\theta_0) \delta \theta+\frac{1}{2}\Theta^{\prime\prime}(\theta_0) \delta \theta^2+\dots
\end{align}
where $R^\prime(r_0)=\left.\frac{d R(r)}{dr}\right|_{r_{0}}$, $\Theta^{\prime\prime}(\theta_0) = \left.\frac{d^2 \Theta(\theta)}{d\theta^2}\right|_{\theta_{0}}$ etc. The stability of the orbits requires the first two terms in these explanations to vanish so that from the equations  (\ref{rad}) and (\ref{tet}), one is led to the following equations 
\begin{align}
\label{freq-eqs}
    &g_{rr} \, \delta \dot r^{2}= \frac{1}{2}R^{\prime\prime} (r_0)\delta r^2\ , \quad\nonumber \\
    &g_{\theta \theta} \, \delta \dot \theta^{2}=\frac{1}{2}\Theta^{\prime\prime}(\theta_0)\delta \theta^2 \, .
\end{align}
Now, taking derivatives of these equations for the invariant time $\lambda$ and relating the invariant time to coordinate time as
$\frac{d}{d\lambda}=\frac{dt}{d\lambda}\frac{d}{dt}=\dot t\frac{d}{dt}$ we find the following harmonic oscillation equations  
\begin{align}
    \frac{d^2}{dt^2}\delta r+ \omega_r^2 \delta r =0\nonumber\  , \quad
    \frac{d^2}{dt^2}\delta \theta+ \omega_\theta^2 \delta \theta =0
\end{align}
wherein the radial and the vertical (lateral)
frequencies have the explicit expressions 
\begin{align}
    \omega_r^2=-\frac{1}{ g_{rr}\dot{t}^2}R^{\prime\prime} (r_0) \ , \quad
    \omega_\theta^2=-\frac{1}{ g_{\theta\theta}\dot{t}^2}\Theta^{\prime\prime}(\theta_0)\ .
\end{align}
After substituting the metric functions, $R^{\prime\prime}(r_0)$ and $\Theta^{\prime\prime}(\theta_0)$, the radial and vertical frequencies, respectively, read as

\begin{align}
    \omega_r^2=&\Bigg(24 \left(3 l^4 q_1^2 (\sigma -1) r^{\sigma +4}
    +(\sigma -3) \left(3 l^2 r^7\right. \right.  \nonumber\\
    &\left.\left.  -l^4 r^4 \left(q_1 q_2 \left(\sigma ^2-4\right)-3 r+6\right)\right)\right)\Bigg)^{-1}\nonumber\\
   &\left(l^2 \left(3 q_1^2 (\sigma -1) r^{\sigma }+2 q_1 q_2 \left(\sigma ^2-7\right)-12 (r-2)\right)-12 r^3\right)\nonumber \\ 
   &\left(l^2 \left(3 q_1^4 (\sigma -1)^3 r^{2 \sigma }+q_1^2 (\sigma -1) r^{\sigma } \left(q_1 q_2(\sigma -2) (\sigma +2)\right.\right.\right. \nonumber \\
    &\left.\left. (\sigma  (\sigma +4)-6)+3 \left(-\left((r-2) \sigma ^2\right)+r+8 \sigma -12\right)\right)\right. \nonumber \\
    &\left. +(\sigma -3) \left(q_1 q_2 \left(\sigma ^2-4\right)+6\right) \left(q_1 q_2 \left(\sigma ^2-4\right)-r+6\right)\right)\nonumber\\
    &\left.-3 r^3 (\sigma -3) \left(q_1^2 (\sigma -5) (\sigma -1) r^{\sigma }-5 q_1 q_2 \left(\sigma ^2-4\right)\right. \right.\nonumber\\
    &\left.\left.+8 r-30\right)\right) \, , 
\end{align}

\begin{align}
    \omega_\theta^2=&\frac{1}{6 l^2 r^3 (\sigma -3)}\left(3 l^2 q_1^2 (\sigma -1)^2 r^{\sigma }\right.\nonumber\\
    &\left.+(\sigma -3) \left(l^2 \left(q_1 q_2 \left(\sigma ^2-4\right)+6\right)+6 r^3\right)\right) \, .
\end{align}

In our analysis, we express all the frequencies in ${\rm Hz}$ namely, we define the epicyclic frequencies 
\begin{eqnarray}
\nu_i = \frac{1}{2\pi}\frac{c^3}{GM}\, \omega_i
\end{eqnarray}
where $i=(r,\theta,\phi)$, $c=3\cdot 10^8\ \rm m/sec$ is the speed of light in vacuum, $G=6.67\cdot 10^{-11}\ \rm m^3/(kg^2\cdot sec)$ is the gravitational Newtonian constant, and $M=10 M_{\odot}$, $M_{\odot}$ being the solar mass.

\begin{figure}[ht!]
 \centering
   \includegraphics[scale=0.58]{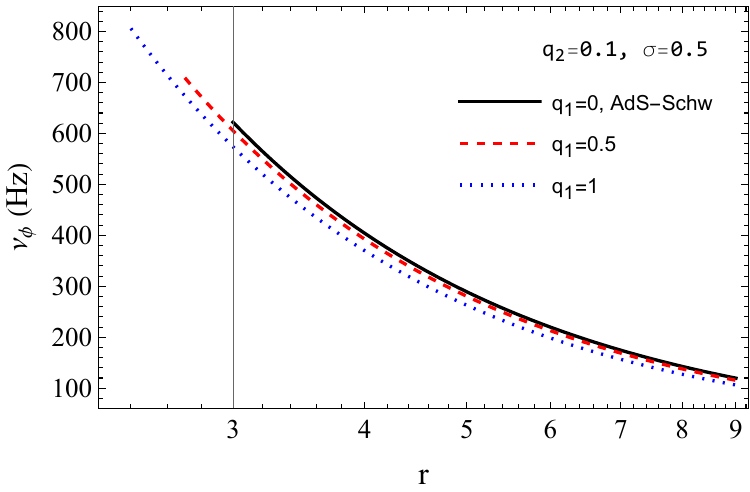}
   \includegraphics[scale=0.58]{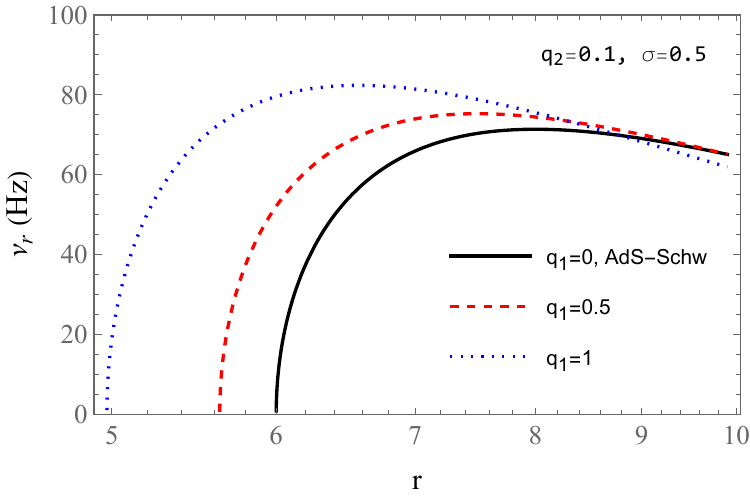}
 \caption{The epicyclic frequencies of the Keplerian (upper panel) and radial (lower panel) types for the indicated values of the EMPG parameters.}\label{hf}
\end{figure}

Frequencies of particles in Keplerian orbits and their radial oscillation frequencies are plotted in Fig. \ref{hf} as a function of the radial coordinate $r$. In the figure, the upper (lower) panel is for the Keplerian (radial) frequency $\nu_\phi$ ($\nu_r$).  The vertical line in the upper panel stands for the photon sphere radius of the Schwarzschild black hole. One notes that the Keplerian frequency and lateral frequencies coincide in the equatorial plane. As the figure shows, for $q_1>0$, the frequencies fall below that of the Schwarzschild black hole. For the radial frequency in the lower panel, the peak moves to higher $r$ values as $q_1$ gets larger. In both cases we recover the radial and lateral frequency of AdS-Schwarzschild solution by considering the limit case of $q_1=0$, which is shown with solid black lines.

\section{QPOs and Its Astrophysical Applications\label{sect:5}}

In observational astrophysics, high-frequency quasi-periodic oscillations (HF QPOs) hold significant importance, serving as a robust framework for predicting the parameters of black holes in accretion systems. These systems include microquasars, binaries with a stellar-mass black hole, and active galactic nuclei housing supermassive black holes. The HF QPO frequencies in microquasars typically fall within the hundreds of ${\rm Hz}$ range, while those around supermassive black holes are significantly lower, differing by orders of magnitude. The observed inverse-mass scaling in these frequencies, governed by the relationships of epicyclic frequencies in orbital motion \cite{Rem-McCli:2005:ARAA:}, makes geodesic models of HF QPOs promising. Particularly noteworthy is their tendency to reveal themselves in rational ratios, often in a 3:2 ratio \cite{Tor-etal:2011:ASTRA:}, suggesting resonant phenomena. In geodesic models \cite{Klu-Abr:2001:ACTAASTR:}, including the electromagnetic interactions \cite{Kol-Tur-Stu:2017:EPJC:},  both of the observed upper frequencies $\nu_\mathrm{u}$ and lower frequencies $\nu_\mathrm{l}$ are theorized to result from a combination of the orbital and epicyclic frequencies, also relevant to slender tori oscillations \cite{Rez-Yos-Zan:2003:MNRAS:}. The geodesic models were introduced initially in the relativistic precession (RP) model, with the identification $\nu_\mathrm{u} = \nu_{\phi}=\nu_\mathrm{K}$ and $\nu_\mathrm{l} = \nu_\mathrm{K} - \nu_\mathrm{r}$ \cite{Ste-Vie:1999:ApJ:}. In this section, we employ epicyclic frequencies in the RP model for twin HF QPOs. 

\begin{figure}[ht!]
\centering
\includegraphics[scale=0.8]{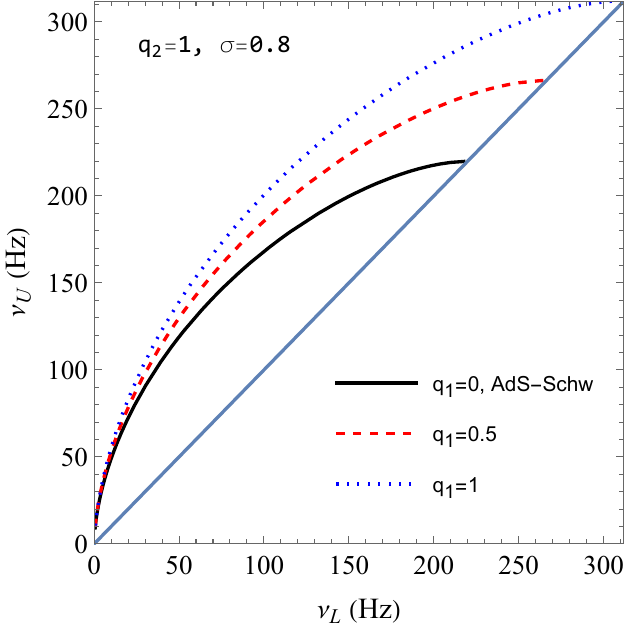}
\includegraphics[scale=0.8]{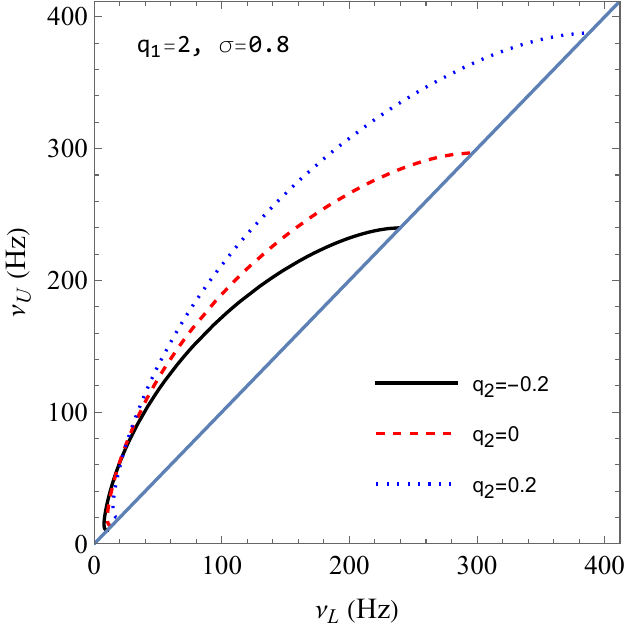} 
 \caption{Correlation between the upper and lower QPO frequencies for the indicated values of parameters $q_1$ and $q_2$. In the upper (lower) panel, $q_2$ ($q_1$) is held fixed, and $q_1$ ($q_2$) is allowed to take different values.}\label{uplow}
\end{figure}
We plot the correlation between the upper ($\nu_U$) and lower ($\nu_L$) frequencies of twin QPO in Fig. \ref{uplow} for different values of $q_1$ and $q_2$. Here, we set $\sigma=0.8$. It is observed that the frequency ratio $\nu_U:\nu_L$ grows with the increase of $q_1$ and $q_2$. The same ratio decreases for negative values of $q_2$.

\section{Constraining EMPG parameters from the twin-peak QPO data\label{sect:6}}

In this section, we determine bounds on model parameters using the twin-peak QPO data from three different types of black hole sources: {\it (i)} stellar mass BH (sMBH), {\it (ii)} super-massive BH (SMBH), {\it (iii)} and intermediate-mass BH (IMBH). These black holes are, respectively, the GRO 1655-40, Sgr-A*, and M82-X1. Their observational data are documented in Table \ref{table1}. We use these results to investigate the three black hole systems mentioned above to constrain the parameters of the EMPG model ({\it i.e.,} the Einstein-Geometric Proca-AdS space-time).

\begin{table*}[ht!]
\begin{center}
      \caption[]{\label{table1} Epicylic frequencies of the twin-peak QPOs in microquasars and galactic center.}
\renewcommand{\arraystretch}{1.2}
\begin{tabular}{| l || c  c | c  c | l |}
            \hline
    Source$^{~\rm (a)\,}$ &  $\nu_{\rm{U}}\,$[Hz]&$\Delta\nu_{\mathrm{U}}\,$[Hz]& $\nu_{\rm {L}}\,$[Hz]&$\Delta\nu_{\rm{L}}\,$[Hz]& Mass$^{~\rm (b)\,}$ [\,M$_{\odot}$\,] \\
\hline
GRO~J1655--40 (sMBH)   & 441&$\pm\,2$& 298&$\pm\,4$ & 5.4$\pm$0.3   \\
\hline
Sgr~A* (SMBH) & $(1.445$ & $\pm 0.16)\times 10^{-3}$  & $(0.886$ &$\pm 0.04)\times 10^{-3}$ & (4.1 $\pm$ 0.6)\,$\times 10^6$\\
\hline
M82-X1  (IMBH) & 3.32&$\pm\,0.06$& 5.07&$\pm\,0.06$&  $415\pm 63$   \\
\hline
\end{tabular}
\begin{list}{}{}
\item[$^{\rm{(a)}}$] Twin-peak QPOs were first reported in Refs.\cite{Strohmayer2001,Ghez:2008,2015MNRAS.451.2575S}.
\item[$^{\rm{(b)}}$] Given in Refs. \cite{Strohmayer2001,Ghez:2008,2015MNRAS.451.2575S}.
\end{list}
\end{center}
\end{table*}

\subsection{Monte Carlo Markov Chain (MCMC) priors for EMPG parameters}
In this subsection, we utilize the well-equipped library \texttt{emcee} \cite{emcee} to perform an MCMC analysis to obtain constraints on the EMPG parameters for a test particle around the compact object. In our implementation, we use the relativistic precision (RP) method.

The posterior distribution can be defined according to the standard definition \cite{emcee,2024PDU....4601561M,2024PhRvD.109j4074A},
\begin{eqnarray}
\mathcal{P}(\theta |\mathcal{D},\mathcal{M})=\frac{P(\mathcal{D}|\theta,\mathcal{M})\pi (\theta|\mathcal{M})}{P(\mathcal{D}|\mathcal{M})}
\end{eqnarray}
where the Gaussian prior for $\theta$ is selected within appropriate limits (as listed in Table \ref{table2}), with $\pi(\theta)$ serving as the prior distribution
\begin{eqnarray}
\pi(\theta_i) \sim \exp\left[{\frac{1}{2}\left(\frac{\theta_i - \theta_{0,i}}{\Tilde{\sigma}_i}\right)^2}\right]
\end{eqnarray}
where $\theta_{\text{low},i} < \theta_i < \theta_{\text{high},i}$.
In our MCMC simulation of the EMPG model, the free parameters are $\theta_i = \{M,q_1,q_2,\sigma,r/M\}$ with their respective standard deviations $\Tilde{\sigma}_i$. 
It is worth mentioning that we keep the black hole mass explicit for this analysis.

Finally, the likelihood function $\mathcal{L}=P(D|\theta,\mathcal{M})$ has the form
\begin{eqnarray}
\log {\cal L} = \log {\cal L}_{\rm U} + \log {\cal L}_{\rm L},\label{likelyhood}
\end{eqnarray}
in which $\log {\cal L}_{\rm U}$ is the log-likelihood of the orbital frequency,
\begin{eqnarray}
 \log {\cal L}_{\rm U} &=& - \frac{1}{2} \sum_{i} \frac{(\nu_{\phi\rm, obs}^i -\nu_{\phi\rm, th}^i)^2}{(\Tilde{\sigma}^i_{\phi,{\rm obs}})^2},
\end{eqnarray}
and $\log {\cal L}_{\rm L}$ stands for the log-likelihood of the periastron precession frequency ($\nu_{\rm per}$) as follows:
\begin{eqnarray}
\log {\cal L}_{\rm L} =-\frac{1}{2} \sum_{i} \frac{(\nu_{\rm per, obs}^i -\nu_{\rm per, th}^i)^2}{(\Tilde{\sigma}^i_{\rm per,{\rm obs}})^2}.
\end{eqnarray}
In the above, $\nu^i_{\phi,\rm obs}$ and $\nu^i_{\rm per,\rm obs}$ are the observed values for the orbital and periastron precession frequencies ($\nu_{\rm K}=\nu_{\rm U}$) for the sources under concern. On the other hand, $\nu^i_{\phi,\rm th}$ and $\nu^i_{\rm per,\rm th}$ stand for the values predicted by the EMPG model (see Sec. \ref{sect:4}). 
The periastron precession frequency $\nu_{\text{per}}$ can be calculated by subtracting the radial frequency $\nu_{\rm r}$ from the Keplerian frequency $\nu_{\rm K}$. 

\begin{table*}[ht!]
\begin{center}
\renewcommand\arraystretch{1.5} 
\caption{\label{table2}%
The Gaussian priors ($\mu$ is the mean value and ${\bar{\sigma}}$ the variance) of the EMPG model from QPOs of the sources sMBH, SMBH, and IMBH.}
\begin{tabular}{lcccccccccc}
\hline\hline
\multirow{2}{*}{Parameters} & \multicolumn{2}{c}{GRO J1655-40} & \multicolumn{2}{c}{Sgr-A{}*} & \multicolumn{2}{c}{M82-X1} \\
                            & $\mu$ & \multicolumn{1}{c}{$\Tilde{\sigma}$} & $\mu$          & $\Tilde{\sigma}$         & $\mu$ & \multicolumn{1}{c}{$\Tilde{\sigma}$}       \\
\hline
     $ M\; (M_{\odot})$ & $5.35$ & 0.12 & $4.261\times 10^6$ & $0.072\times 10^6$ & $417.5$ & 5.27\\
     $|q_1|$ & 1.25 &0.15   & 1.37  & 0.10 & $1.25$ & 0.11 \\
     $q_2$ & 0.18 &0.07 & 0.13 &0.09 & 0.25 & 0.08 \\
     $\sigma$ & $0.71$ & 0.09 & $0.61$ & 0.09 & $0.61$ & 0.11\\
     $r/M$ & $5.76$ & 0.15 & $5.60$ & 0.16 & $5.46$ & 0.16\\
     \hline\hline
\end{tabular}
\end{center}
\end{table*}

To establish the priors, we begin by examining the minimization of the Chi-square for Eq. (41) while setting $l=10^{12}$, $q_1>1$, and $\sigma>0.5$, where the Proca parameters have a significant effect. We explore the parameter spaces outlined in \cite{AdS-Proca-1} to do this. This allows us to determine the values of the remaining parameters, $M$, $q_2$, and $r/M$, that result in the minimum likelihood function and ensure that the mass of the compact source closely matches the observation. This process leads us to identify the appropriate priors.

Having set the priors, we use the available data to perform an MCMC simulation to determine the plausible range of the EMPG parameters  \{$M$, $q_1,q_2,\sigma$, $r/M$\} for Einstein-Geometric Proca AdS compact object. Considering the Gaussian prior distribution, we sample $10^5$ points for every parameter. This approach allows us to investigate the physically allowed multi-dimensional parameter space within defined limits and obtain the parameter values that best match the data.

\begin{figure*}[ht!]
    \centering
    \includegraphics[height=10cm,width=10cm]{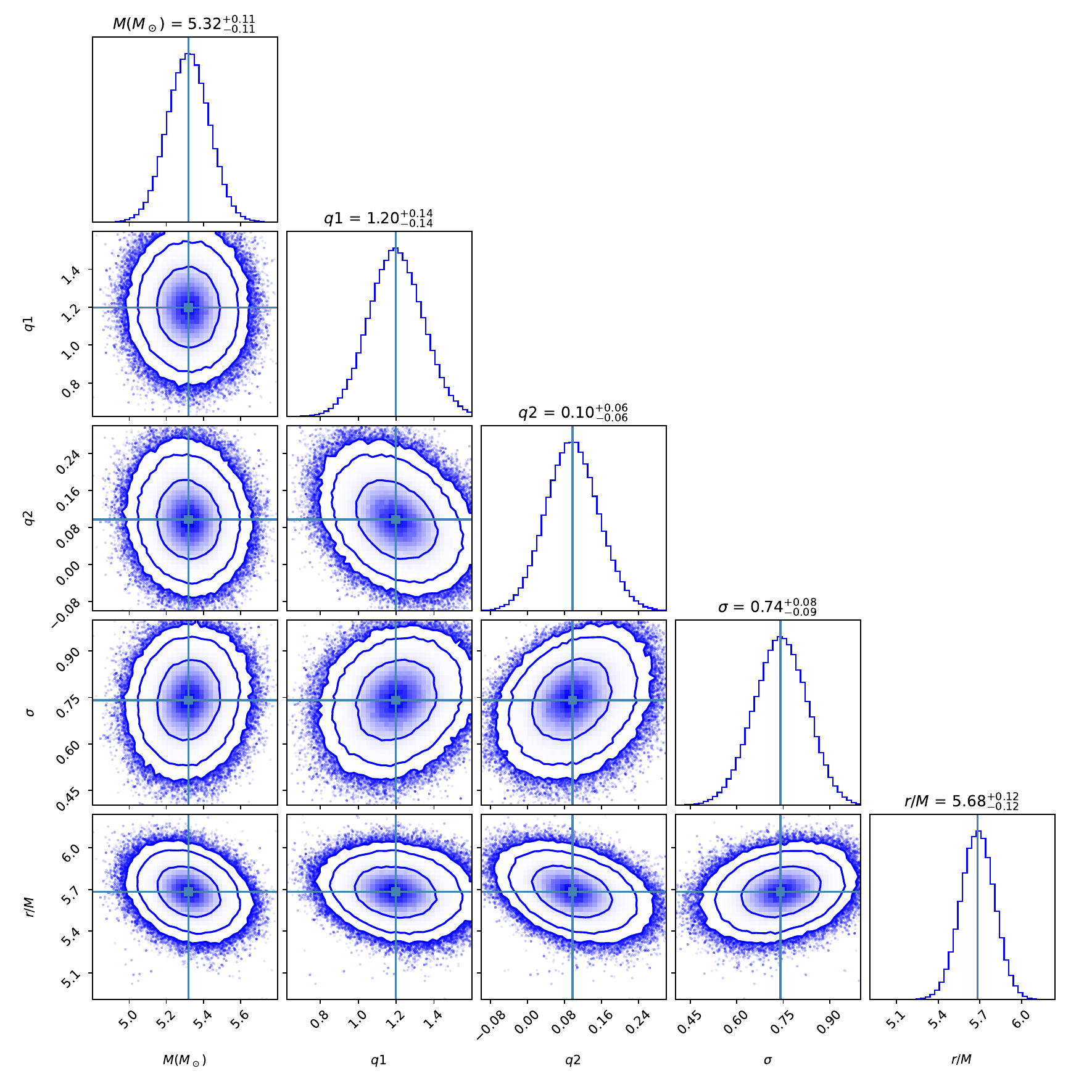}
    \quad
    \includegraphics[height=10cm,width=10cm]{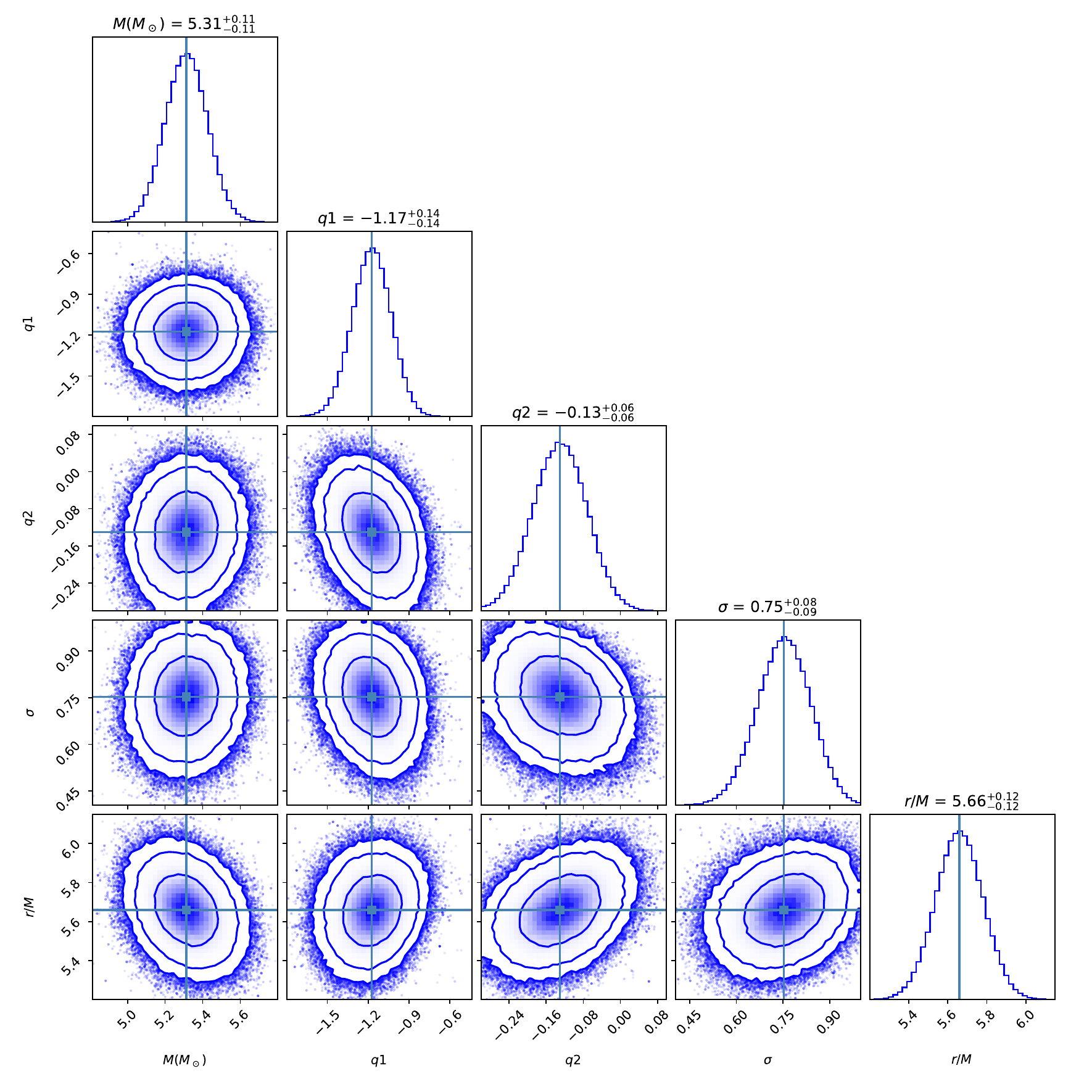}
    \caption{Constraints on the EMPG model parameter from a five-dimensional MCMC simulation using the state-of-the-art QPO data for the stellar-mass black hole GRO J1655-40 in the RP model. The blue contours in the upper (lower) panel correspond to $q_1>0$ ($q_1<0$).}
    \label{contour1}
\end{figure*}

\begin{figure*}[ht!]
    \centering
    \includegraphics[height=11cm,width=11cm]{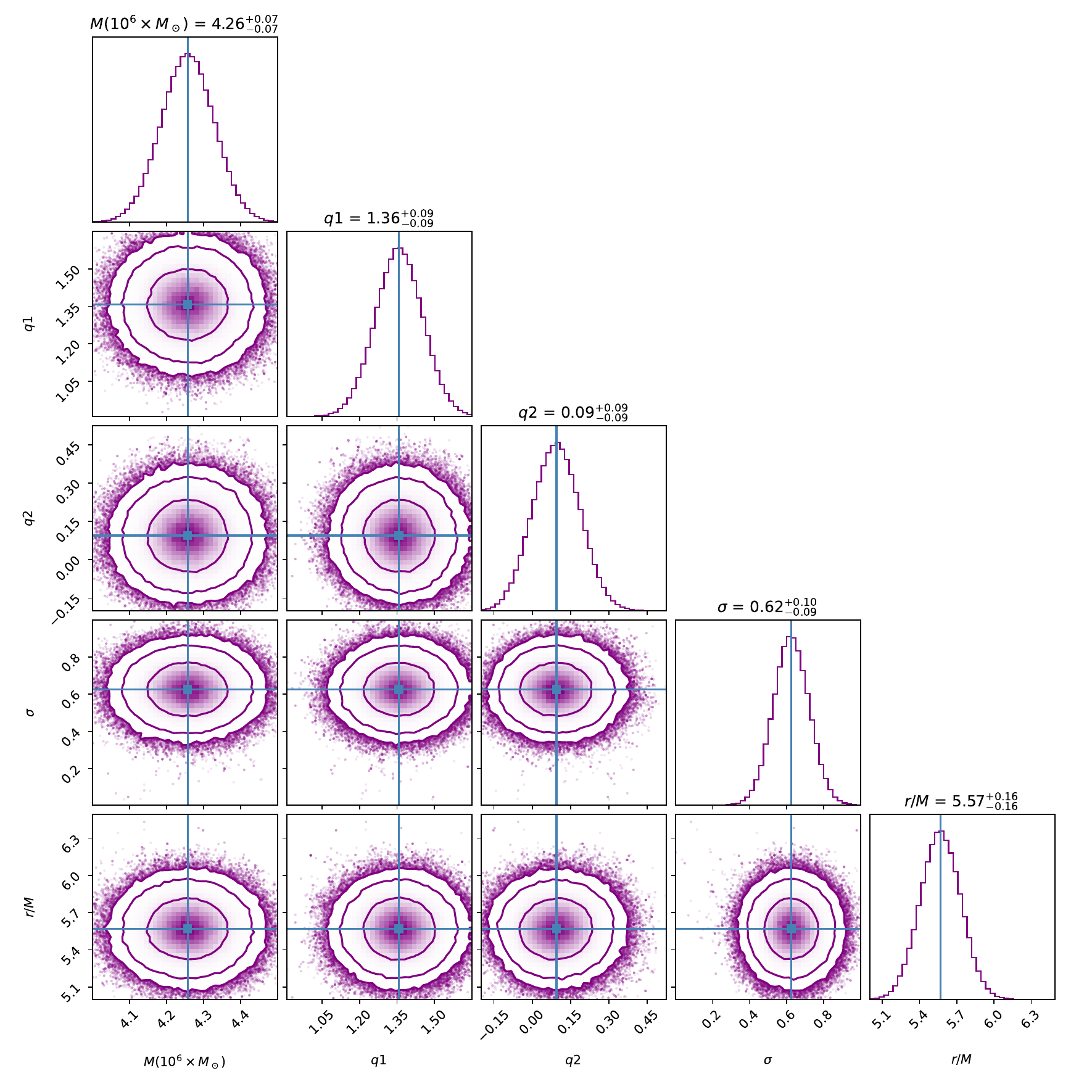} \quad 
    \includegraphics[height=11cm,width=11cm]{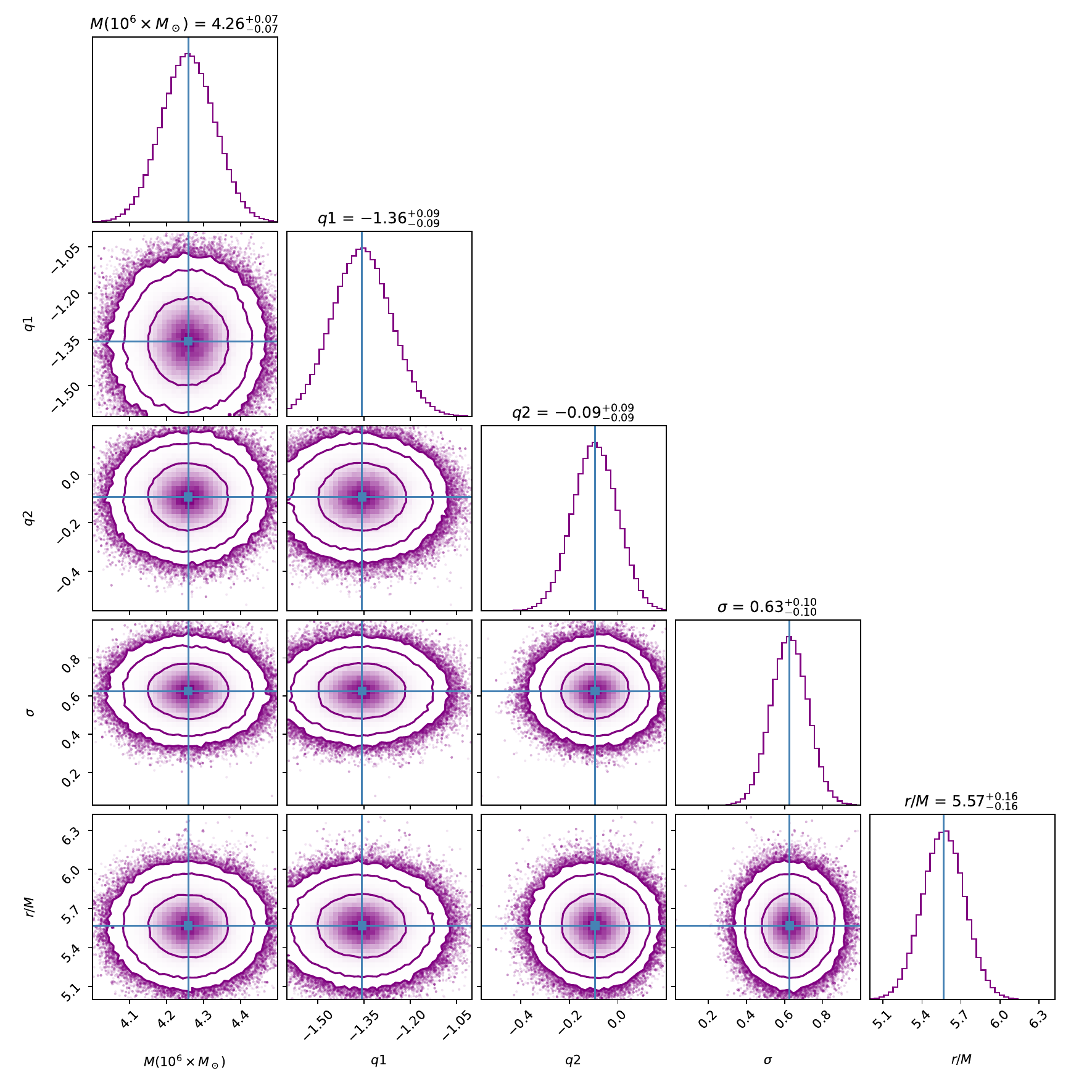} \caption{The same as in Fig. \ref{contour1} but for the supermassive black hole Sgr-A{}*.}
    \label{contour2}
\end{figure*}

\begin{figure*}[ht!]
    \centering
    \includegraphics[height=11cm,width=11cm]{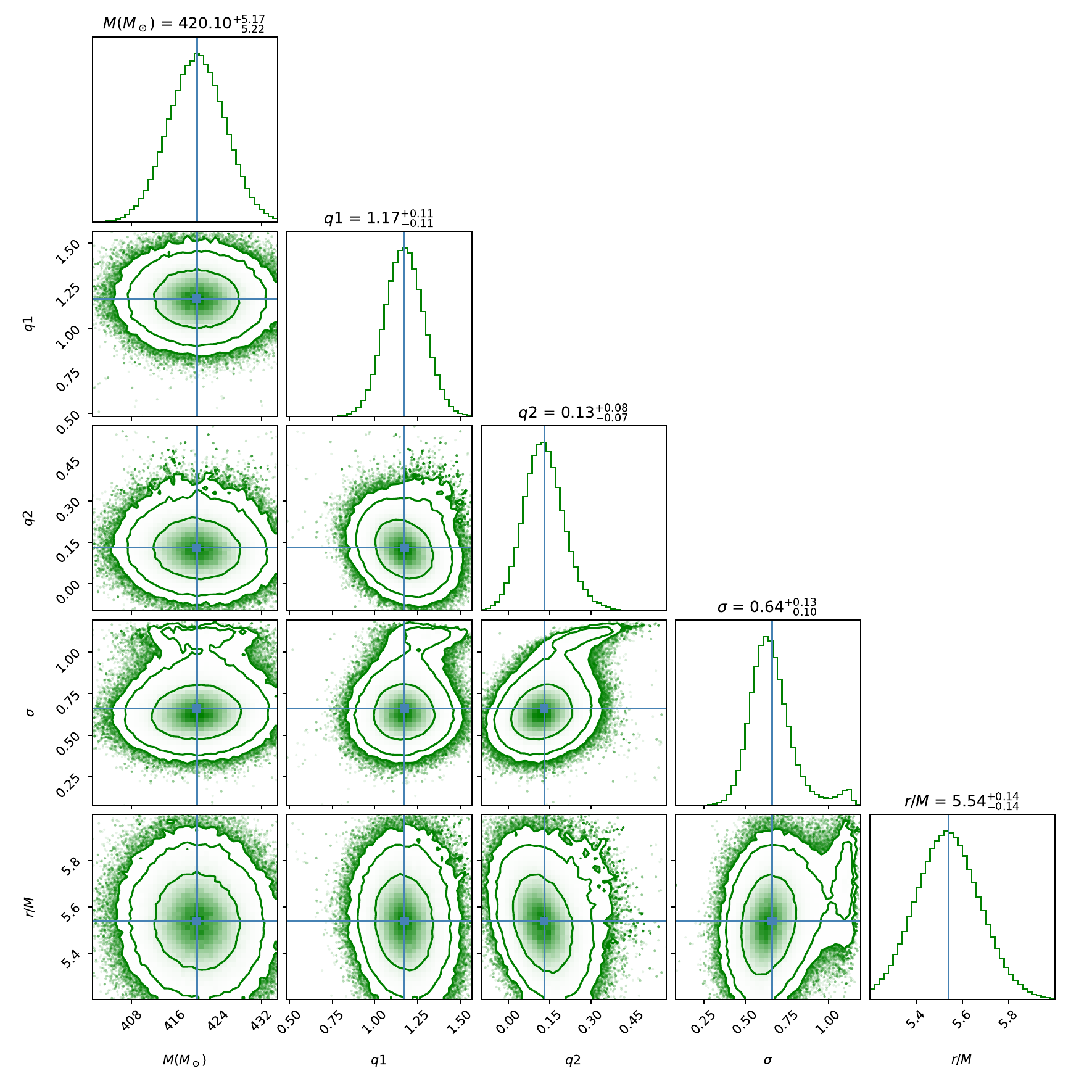} \quad \includegraphics[height=11cm,width=11cm]{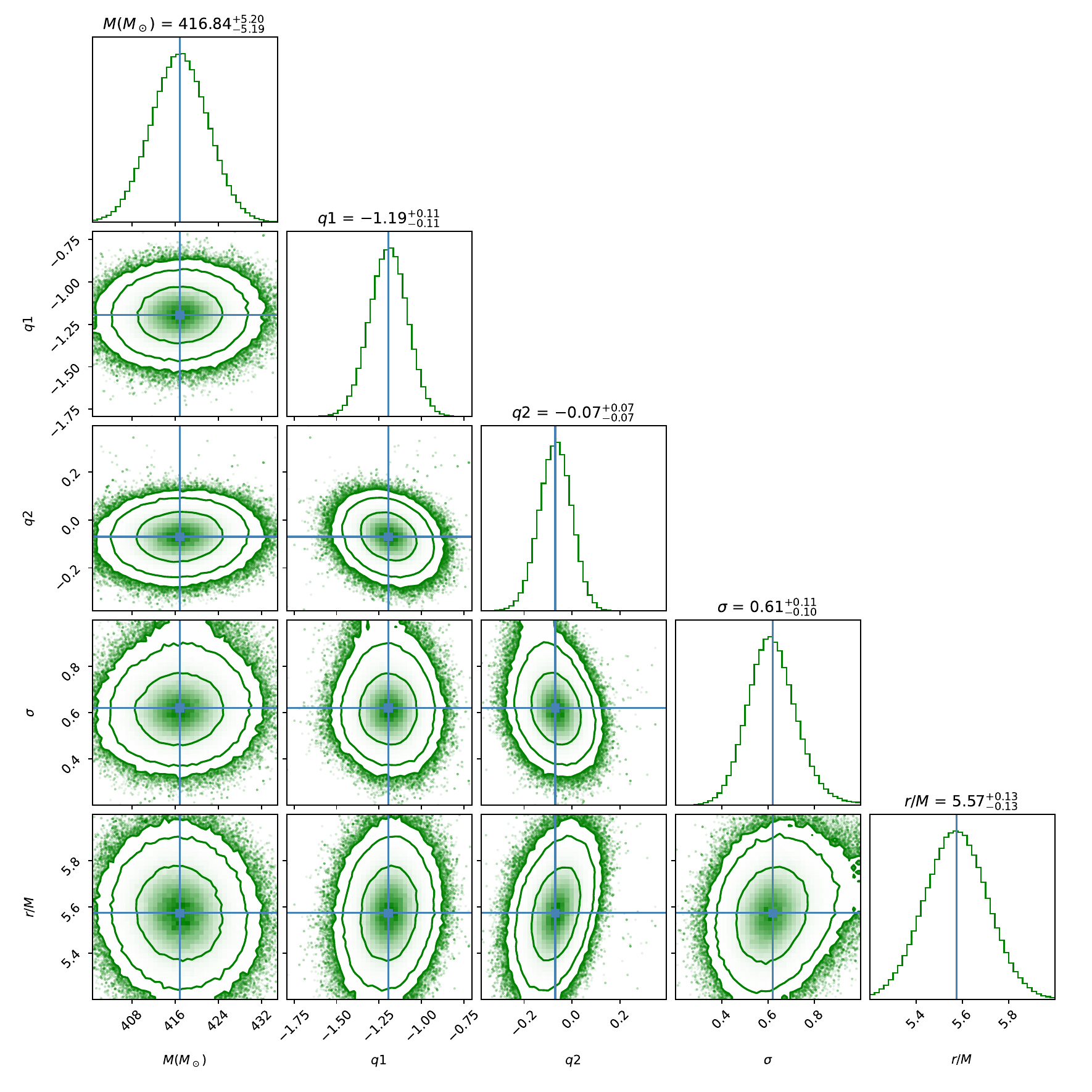}
    \caption{The same as in Fig. \ref{contour1} but for the intermediate-mass black hole M82-X1.}
    \label{contour3}
\end{figure*}

\subsection{Results of the MCMC simulation}
Using the setup described above, we investigated the 5-dimensional EMPG parameter space via an MCMC analysis for a test particle in the Einstein-Geometric Proca AdS spacetime. Note that in the MCMC simulation, we set the AdS radius parameter $l=10^{12}$ unless otherwise noted. This way, we constrain the mass of compact objects ($M$), the ranges of $q_1$ and $q_2$, the mass parameter $\sigma$, and the epicyclic oscillation radius ($r/M$), respectively.

With this, in Fig. \ref{contour1}, Fig. \ref{contour2}, and Fig. \ref{contour3}, we give the results of the MCMC analysis of the EMPG model parameters by using the data from three different astrophysical sources. Using the \textit{emcee} package, we obtain the posterior distribution with the possible choice of the prior given in Table \ref{table1}. In doing so, we choose $q_1$ to be both positive and negative. With the higher values of $|q_1|>1$, we notice that the mass-parameter converges to $\sigma > 0.6$, which suggests the influence of the EPMG parameter on the QPOs. However, both the choice of prior and the posterior distribution suggest the strength of $q_2$ to be smaller. Interestingly, for higher $\sigma$ and $q_1$, the radius of epicyclic oscillation is converging towards $r/M < 6$, consistent with the Fig. \ref{iscoplot}. In all the MCMC analyses, the contour plots highlight the confidence levels (1$\Tilde{\sigma}$ ($68\%$), 2$\Tilde{\sigma}$ ($95\%$) and 3$\Tilde{\sigma}$ ($99\%$) of the posterior probability distributions for the entire set of parameters. The shaded regions on the contour plots reflect these confidence levels.

\begin{table*}[ht!]
\begin{center}
\renewcommand\arraystretch{1.5} 
\caption{\label{table3}%
 The best-fit posterior distributions of the EMPG parameters from the QPO data.}
\begin{tabular}{|l|c|c|c|l|}
\hline\hline
\multirow{1}{*}{Parameters} & \multirow{1}{*}{GRO J1655-40} & \multirow{1}{*}{Sgr-A{}*} & \multirow{1}{*}{M82-X1} & \multirow{1}{*}{Remark} \\
\hline
     $ M\; (M_{\odot})$ & $5.32^{+0.11}_{-0.11}$ & $4.26^{+0.07}_{-0.07} \times 10^6$ & $420.10^{+5.17}_{-5.22}$ \\
     $q_1$ & $1.20^{+0.14}_{-0.14}$    & $1.36^{+0.09}_{-0.08}$  & $1.17^{+0.11}_{-0.11}$ & $q_1>0$\\
     $q_2$ & $0.10^{+0.06}_{-0.06}$ & $0.09^{+0.09}_{-0.09}$ & $0.13^{+0.08}_{-0.07}$  \\
     $\sigma$ & $0.74^{+0.08}_{-0.09}$ & $0.62^{+0.10}_{-0.09}$ & $0.64^{+0.13}_{-0.10}$ \\
     $r/M$ & $5.68^{+0.12}_{-0.12}$ & $5.57^{+0.16}_{-0.16}$ & $5.54^{+0.14}_{-0.14}$ \\

     \hline
     $ M\; (M_{\odot})$ & $5.31^{+0.11}_{-0.11}$ & $4.26^{+0.07}_{-0.07} \times 10^6$ & $416.84^{+5.20}_{-5.19}$ \\
     $q_1$ & $-1.17^{+0.14}_{-0.14}$    & $-1.36^{+0.09}_{-0.09}$  & $-1.19^{+0.11}_{-0.11}$  & $q_1<0$\\
     $q_2$ & $-0.13^{+0.06}_{-0.06}$ & $-0.09^{+0.09}_{-0.09}$ & $-0.07^{+0.07}_{-0.07}$ \\
     $\sigma$ & $0.75^{+0.08}_{-0.09}$ & $0.63^{+0.10}_{-0.10}$ & $0.61^{+0.11}_{-0.10}$ \\
     $r/M$ & $5.66^{+0.12}_{-0.12}$ & $5.57^{+0.16}_{-0.16}$ & $5.57^{+0.13}_{-0.13}$ \\
     \hline


\end{tabular}
\end{center}
\end{table*}

More information on the best-fit values of these three parameters is given in Tab.~\ref{table3}.

However, it is worth noting that when $q_1 = 0$, the space-time simplifies to the well-known Schwarzschild AdS (Sch-AdS) space-time, which is only parametrized by $l$. We have incorporated these details in our MCMC code.


\section{Conclusion\label{sect:7}}
In this paper, we have constructed spherically symmetric solutions of the extended Einstein-Geometric Proca-AdS (EMPG) model \cite{Demir2020,dp-yeni,AdS-Proca-1} and put constraints on its parameters using the observational data on QPOs of stellar mass black holes, supermassive black holes, and intermediate-mass black holes. The geometric nature of the Proca field is a distinguishing feature compared to the generic Einstein-Proca systems and the $Z^\prime$ gauge boson in the literature. 

Our investigation of the EMPG involves its static, spherically symmetric solutions and the circular orbits of test particles around them. Our analysis started with deriving the effective potential and determining the energy and angular momenta from it.  We have shown that when $q_1=q_2$, the angular momentum attains its critical values (see the circled dots in Fig\ref{cr}). As $q_1$ increases, the critical values of angular momentum and energy become smaller. However, when $q_1$ and $q_2$ are fixed, the critical angular momentum takes a larger value with the AdS mass parameter $\sigma$ growth. We have also studied the ISCOs of the particles. When one is kept constant, the ISCO radius decreases with an increase in $q_1$ or $q_2$. However, it gets larger as $\sigma$ approaches to 1. At $q_1=0$, the ISCO radius remains constant for all values of $q_1$ \& $q_2$ and nearly equals the ISCO radius in the Schwarzschild case, as anticipated. Moreover, in the case of $q_1=0$, the particle's angular momentum at the ISCO reaches its maximum value. At the same time, the energy attains its minimum value, aligning with the characteristic values in the Schwarzschild black hole having $r_{ISCO}=6$. Also, as $q_1$ increases, the range for the ISCO radius expands, giving rise to a broader range of energy and angular momentum. At $q_1\neq 0$, the ISCO angular momentum increases with $\sigma \to 1$. However, in $q_1\neq 0$, the particle experiences its minimum energy in the ISCO, which matches the Schwarzschild case for $\sigma=0$ and $\sigma=1$. The peak in the energy values shifts to higher values of $\sigma$ as $q_1$ increases.

We have also derived expressions for the frequencies of test particles' oscillations in the radial and vertical directions along the circular stable orbits around the EMPG compact object. Our numerical analyses and graphical results have shown that the frequencies get smaller for $q_1>0$ than those in the Schwarzschild case. The radial and peak frequencies increase at higher values of $q_1$. We studied the twin-peak QPOs in the RP model with the frequencies at hand. We have determined the relationships between the upper and lower frequencies in the twin QPOs, fixing, $\sigma=0.8$ and observed that the frequency ratio $\nu_U:\nu_L$ gets enhanced with the increase of $q_1$ \& $q_2$. In contrast, the ratio decreases for the cases $q_2<0$. 

Finally, we have performed a detailed MCMC analysis of the EMPG model parameters using the observational data from the twin-peak QPOs observed in the stellar-mass black hole candidate in GRO J1655-40, the intermediate-mass black hole in M82-X1, and the supermassive black hole SgA* in our Milky Way. In the MCMC analysis, we searched for the best-fit values of the EMPG parameters $M$, $q_1,q_2,\sigma$ and $r/M$ (radius of the QPO orbit) using the QPOs observed in the objects mentioned above. We have divided our analysis into two parts: positive values of $q_1>0$ and negative values of $q_1<0$ as presented in Figs. \ref{contour1},  \ref{contour2}, and \ref{contour3}. The shaded regions on these contour plots stand for the confidence levels.

We have performed a detailed study of metric-Palatini gravity extended with the antisymmetric part of the affine curvature. Our analyses show that with the increasing precision in astrophysical observations, it will be possible to pin down the model parameters like $\sigma$, $q_1$, and $q_2$ to a satisfactory level. Our solution and its phenomenological analysis can be regarded as part of a general program of constraining metric-affine gravity theories using the black hole and other observations. 

\section*{Acknowledgements}
Dedicated to Durmu\c{s} Demir (1967-2024), our supervisor and candid friend.
This research is partly supported by Grant F-FA-2021-510 of the Uzbekistan Ministry for Innovative Development.  D. D. and B. P.  acknowledge the contribution of the COST Action CA21106 - COSMIC WISPers in the Dark Universe: Theory, astrophysics, and experiments (CosmicWISPers). The work of B. P. is supported by the Astrophysics Research Center of the Open University of Israel (ARCO) through The Israeli Ministry of Regional Cooperation. The work of B. P. is supported by Sabanc{\i} University, Faculty of Engineering and Natural Sciences.

\bibliographystyle{spphys}
\bibliography{reference}


\end{document}